**Active Cytoskeletal Composites Display Emergent Tunable Contractility and Restructuring**

Gloria Lee[1], Gregor Leech[1], Pancy Lwin[2], Jonathan Michel[2], Christopher Currie[1], Michael J. Rust[3], Jennifer L. Ross[4], Ryan J. McGorty[1], Moumita Das[2], and Rae M. Robertson-Anderson[1]

[1]Department of Physics and Biophysics, University of San Diego
[2]School of Physics and Astronomy, Rochester Institute of Technology
[3]Department of Molecular Genetics and Cell Biology, University of Chicago
[4]Department of Physics, Syracuse University

The cytoskeleton is a model active matter system that controls diverse cellular processes from division to motility. While both active actomyosin dynamics and actin-microtubule interactions are key to the cytoskeleton's versatility and adaptability, an understanding of their interplay is lacking. Here, we couple microscale experiments with mechanistic modeling to elucidate how connectivity, rigidity, and force-generation affect emergent material properties in in vitro composites of actin, tubulin, and myosin. We use time-resolved differential dynamic microscopy and spatial image autocorrelation to show that ballistic contraction occurs in composites with sufficient flexibility and motor density, but that a critical fraction of microtubules is necessary to sustain controlled dynamics. Our active double-network models reveal that percolated actomyosin networks are essential for contraction, but that networks with comparable actin and microtubule densities can uniquely resist mechanical stresses while simultaneously supporting substantial restructuring. Our findings provide a much-needed blueprint for designing cytoskeleton-inspired materials that couple tunability with resilience and adaptability.

The cytoskeleton is an active network of proteins, including semiflexible actin filaments, rigid microtubules, and motor proteins, that maintains cell structure and facilitates essential functions such as migration[1,2], division[3–5], and intracellular transport[6–8]. These diverse processes are driven by the cytoskeleton's ability to actively self-organize and restructure, for example by myosin II motors actively crosslinking, pushing and pulling actin filaments[9,10]. Actomyosin interactions generate contractile forces and flow[11], playing an important role in processes ranging from contractile ring formation during cell division[12] to lamellipodia adhesion during cell migration[13].

In vitro studies of actomyosin networks show that active material properties such as contraction can only be achieved by fine-tuning the densities of actin, passive crosslinkers, and myosin[14–20]. In crosslinked actin networks, contraction only occurs for a limited range of crosslinker:actin ratios: too low (<0.05) and the network is insufficiently connected to transmit force, while too high (>0.15) and the network becomes too rigid to respond[16]. Within the active range, reported contraction rates varied by a factor of ~3. Other studies have shown that a critical motor density is required for contractility, but contraction rates do not change substantially above this density[17]. Further, organized contractile dynamics and restructuring has often required confining actomyosin networks to quasi-2D geometries crowded to a surface, and has typically been limited to ~5 min of observed activity[19,20]. Here, to increase the tunability, longevity, resilience, and viability of actomyosin active matter as a functional material, we couple microtubules to actomyosin networks and combine experiments with modeling to systematically explore a broad parameter space of composite formulations and spatiotemporal scales. Our results also shed important new light on the role that actin-microtubule interactions play in actomyosin dynamics in the cytoskeleton.

Steric actin-microtubule interactions are integral to in vivo cytoskeleton processes[21]. The ability of stiff microtubules, with ~100x larger persistence length than semiflexible actin filaments (~1 mm vs ~10 μm) to bear large compressive loads plays an important role in balancing actomyosin contraction during cell migration[22,23]. Meanwhile, actin filaments reinforce the mechanical stability of microtubules by increasing elasticity and reducing their likelihood to buckle[24,25]. Further, previous in vitro studies show that co-entangled actin-microtubule systems exhibit emergent mechanical properties, such as coupled stiffness and mobility, that are not possible in single component systems[26–28], highlighting the promise of cytoskeleton composites as a model platform for material design and engineering.

While active actomyosin systems have been extensively studied, active actin-microtubule composites remain almost entirely unexplored. We previously showed that a co-entangled composite of microtubules and actin filaments driven by myosin II[29] exhibited surprisingly controlled and reproducible contraction in the absence of crosslinkers, which we attributed to microtubules increasing network connectivity and providing extensile reinforcement against flow[30,31]. Here, to build on this proof-of-concept study, we couple time-resolved differential dynamic microscopy and spatial image autocorrelation with active double-network modeling to map the phase space of active dynamics to composite formulation. We reveal how the myosin-driven dynamics of varying composites evolve over long times, and connect the active dynamics to network restructuring and mechanical moduli.

**Results**

We create and simulate co-entangled actin-microtubule networks driven by myosin II motor activity and systematically tune the molar actin fraction ($\Phi_A$=[actin]/([actin]+[tubulin])) and molar myosin concentration ($c_M$) (Fig. 1). We image composites using multi-spectral confocal microscopy (Movie S1) and employ differential dynamic microscopy (DDM) and spatial image autocorrelation (SIA) to quantify the active dynamics as well as the spatiotemporally varying composite structure. Further, we perform simulations of active biopolymer double-networks to understand the mechanisms underlying our experimental results and correlate measured dynamics and restructuring to mechanical properties. To determine the fraction of actin necessary to maintain activity and the fraction of microtubules necessary to maintain controlled dynamics, we vary $\Phi_A$ while holding the combined concentration of actin and tubulin fixed. To determine what range of myosin concentrations drives organized contractility over sustained periods of time, we vary the molar myosin concentration $c_M$ from 0.12-0.48 µM and examine network dynamics over extended time periods (~45 mins to several hours). We note that the protein concentrations used here are lower than those found in cells, with reported values of ~50-500, ~5-20, and ~0.5-5 µM for actin, tubulin and myosin II, respectively[32–34]. We instead honed our parameter space through ample trial-and-error testing based on previous in vitro studies of actomyosin and actin-microtubule systems that showed potential as tunable, resilient, active materials. We show that increasing molar actin fraction $\Phi_A$ and myosin concentration $c_M$ generally results in greater contraction and restructuring in both experiments and simulations (Fig. 1). However, there are important exceptions to this rule and emergent regions of the phase space that we describe below.

To quantify active dynamics and restructuring, we perform differential dynamic microscopy (DDM) on the separate actin and microtubule channels of each time-series collected for each composition. With DDM, we determine the rate at which density fluctuations decay over time[35] (see Methods). Fig. 2A shows representative DDM image structure functions $D(q, \Delta t)$ for the composite formulations shown in Fig. 1. As shown, the dependence of $D$ on the lag time $\Delta t$ is similar for both actin and microtubule channels for each ($\Phi_A$, $c_M$) combination, indicating synchronized movement of actin filaments and microtubules across all composites. However, the lag time at which each $D(q, \Delta t)$ reaches a decorrelation plateau, signifying the time at which the composite is sufficiently decorrelated from its previous configuration, depends strongly on composition.

$D(q, \Delta t)$ curves for networks containing $\Phi_A < 0.5$ and $c_M < 0.12$ µM do not exhibit plateaus within the experimental time frame, indicating undetectable movement. Data taken in the absence of myosin or blebbistatin deactivation show similar curves that lack plateaus (Fig. S1), verifying that the plateaus observed in the other composites arise from motor-driven dynamics (Fig. 2). Of the composites that exhibit plateaus, the lag time at which a plateau is reached markedly decreases with increasing actin fraction. Myosin concentration appears to play a less important role, though a critical $c_M$ is required for detectable active motion.

For image structure functions that reach plateaus, we extract a characteristic decay time $\tau$ for each wavevector $q$ (see Methods) (Fig. 2B). For all cases, $\tau(q)$ curves for actin and microtubule channels overlap, again indicating coordinated dynamics. Further, all curves approximately scale as $\tau \sim q^{-1}$ (Table S1), signifying ballistic motion. The magnitude of $\tau(q)$ determines the contraction rate and depends on composite formulation. By fitting each curve to $\tau \sim (kq)^{-1}$, we determine a composite-dependent contraction velocity, $k$ (Fig. 2C). We find similar contraction velocities in both actin and microtubule channels for all compositions, as expected from similar $\tau(q)$ curves. The composite with ($\Phi_A, c_M$) = (0.75, 0.24) contracts the fastest at 35 ± 3 nm/s, which is ~3x slower than the velocity of 90 ± 20 nm/s we measured for a 100% actin network with the same myosin concentration (Fig. S2). Lowering the actin fraction from $\Phi_A$=0.75 to 0.5 reduces the contraction speed by another factor of ~3 to ~12 nm/s, which is surprisingly unaffected by doubling the myosin concentration from $c_M$=0.24 to 0.48 µM. Further lowering $\Phi_A$ or lowering $c_M$ results in ~6x reductions in contraction velocity, with the two composites exhibiting similar speeds (~2.3 nm/s).

To correlate contraction dynamics with motor-driven restructuring, we analyze the spatial autocorrelation of images at uniformly distributed time-points throughout each time-series (see Methods) (Fig. 3). For a specific image in a time-series, the autocorrelation $g(r)$ determines the degree to which the intensity at one location in the image correlates with the intensity of the surrounding points at varying distances[36]. The more quickly $g(r)$ decays, the smaller the structural features of the network.

As shown in Fig. 3A, $g(r)$ curves generally broaden from initial to final time-points, indicating that structural feature sizes increase during activity. This broadening is particularly noticeable for the $(\Phi_A,c_M)$=(0.75,0.24) composite, where we observe the formation of dense foci separated by large voids during activity (Fig 1A). In general, the slower decay of $g(r)$ suggests that myosin II motors rearrange the composites from homogenous meshes of individual filaments to networks of denser bundles or aggregates separated by larger, sparsely populated gaps. In order to quantify the degree of restructuring and its dependence on composite formulation, we fit $g(r)$ to an exponential $g(r) = Ae^{-\frac{r}{\lambda}}$, where $\lambda$ is the characteristic decay length or correlation length (Fig 3A inset). For an isotropic network, $\lambda$ scales as the mesh size.

Fig 3B shows the average correlation lengths as a function of time, $\lambda(t)$, for actin and microtubule channels for each composition. While the time-dependence of $\lambda$ for both filament types displays a complex dependence on network composition, we can nonetheless crudely separate networks into those that appreciably restructure over time and those that do not. Specifically, we find that $\lambda(t)$ stays roughly constant over time for both $\Phi_A$=0.25 conditions, as well as for $(\Phi_A,c_M)$ = (0.50,0.12). This finding suggests that $\Phi_A$=0.25 networks are too rigid, and $c_M$=0.12 µM networks contain insufficient motor density, to allow for noticeable rearrangement.

While the actin and microtubule networks in all compositions start with approximately the same $\lambda$ values, networks containing the highest actin fraction or highest myosin concentration exhibit the greatest increase in $\lambda$ over time, indicating the greatest degree of restructuring. Interestingly, unlike most other conditions, correlation lengths for actin filaments and microtubules markedly decouple from one another for the $(\Phi_A, c_M) = (0.75, 0.24)$ formulation. While $\lambda(t)$ for the microtubule channel reaches $15 \pm 2$ µm over 45 min, the actin correlation length only increases to $6.4 \pm 0.9$ µm. This effect is also apparent, though subtler, in the $\Phi_A = 0.75$ composite with lower motor concentration $c_M = 0.12$ µM. Given that myosin acts solely on the actin network, the more substantial restructuring of microtubules compared to actin in this composite is particularly noteworthy. Conversely, for the other formulation that exhibits the most extreme restructuring, $(\Phi_A, c_M) = (0.5, 0.48)$, the correlation lengths for actin and microtubule channels both increase ~5x.

To further shed light on these intriguing restructuring data, we generate spatially-resolved contour maps of the contractile and extensile strains experienced by our simulated double-networks (Fig. 4A). As in our experiments, the strain fields for actin and microtubule networks appear correlated for most conditions. While there are slightly more extensile regions in the microtubule network and more contractile regions for actin due to the larger stiffness of microtubules compared to actin, the degree and affinity of deformations are similar for both networks. The striking exception to this rule is the $\Phi_A = 0.75$ composites. Similar to our SIA analysis (Fig. 3B), we see dramatic changes in the structure of the microtubule networks that are distinct from the actin contractility. While the actin network undergoes homogeneous contractility, which is stronger than the other network formulations (more red regions), the microtubule network separates into large distinct regions of deformation.

We also compute the modulus $G'$ of our simulated double-network and show that for a given $\Phi_A$, $G'$ can be significantly enhanced by increasing the concentration of myosin which rigidifies the actin network via contraction into denser meshes and bundles (Fig. 4B). For a given myosin concentration, composites with a dense actin network ($\Phi_A = 0.75$) or microtubule network ($\Phi_A = 0.25$) have a substantially larger $G'$ than for comparable actin and microtubule fractions ($\Phi_A = 0.5$) (Fig. 4). The rigidity in the actin-dense networks arises from more pronounced contraction, while in the microtubule-rich composites, the rigidity comes from the steady-state rigidity of the microtubules themselves. The emergent low-rigidity regime at comparable actin and microtubule concentrations results from both networks being close to the rigidity percolation threshold, thereby allowing for restructuring via non-affine deformations.

To determine whether the varying degrees of composite restructuring and rigidity impact the dynamics over the course of activity, we perform time-resolved DDM over the same 6-min time intervals that we use for our SIA analysis (Fig. 5). Over these shorter time intervals, $D(q, \Delta t)$ curves only reach plateaus for the three cases in which $\Phi_A \geq 0.5$ and $c_M \geq 0.24$ µM. Fig. 5A shows time-resolved $\tau(q)$ curves for these three formulations. For all cases and time-points, $\tau(q)$ curves for actin and microtubules are comparable, similar to our analysis over the entire activity window but at odds with our structural analysis for the $\Phi_A=0.75$ composite. Further, $\tau(q)$ curves for both $\Phi_A = 0.5$ cases exhibit only modest time-dependence compared to the $\Phi_A=0.75$ case which appears to appreciably decrease over the experimental window. By fitting each $\tau(q)$ curve to $\tau \sim (kq)^{-1}$, we compute corresponding time-dependent contraction velocities (Fig. 5B). While the velocities for all composites stay nearly constant for the first ~15 mins, with slight decrease, networks composed of equimolar actin and tubulin ($\Phi_A=0.50$) show minimal change beyond this time while the high $\Phi_A$ case steadily accelerates for the remainder of the activity time.

**Discussion**

To bring together the results of our experimental and theoretical analyses, we construct a cartoon phase diagram that maps measured, simulated, and extrapolated composite characteristics to composite formulation (Fig. 6A). The values provided in the lower panels (Fig. 6B) come directly from experiment or simulation, while the color gradients are qualitative representations of the measured and expected values. The region of formulation space with comparable actin and microtubule concentrations ($\Phi_A=0.5$) and sufficient myosin concentrations ($c_M \geq 0.24$ µM) exhibit measurable contractile dynamics and substantial restructuring without deleterious acceleration or decorrelation of actin and microtubule networks. Additionally, the degree of restructuring and rigidity in $\Phi_A=0.5$ composites can be tuned by increasing myosin concentration without appreciably altering the contraction rate. As such, sustained ballistic contraction and structural integrity over extended periods of time is achieved when composites reach a careful balance between increasing concentrations of motors ($c_M$) and microtubules (1-$\Phi_A$). We discuss this emergent behavior, and other intriguing results, below.

We find that contraction rates depend most strongly on actin fraction rather than myosin concentration, which we attribute to the decreasing composite rigidity that comes with increasing $\Phi_A$. Higher ratios of semiflexible actin filaments to rigid microtubules offer less resistance to motor-generated forces and provide more pathways for force transmission. However, increasing flexibility past a certain threshold compromises material integrity, as evidenced by the accelerating dynamics and decoupling of actin and microtubule networks in the $(\Phi_A, c_M) = (0.75, 0.24)$ case. Our simulations support this interpretation as we find that higher actin fractions result in larger contraction and concomitantly higher modulus $G'$ *following contraction*. Here, $G'$ is a proxy for the spatial range of contraction – a composite with a system-spanning active flexible network and dilute rigid network allows for more and faster restructuring.

While sufficient flexibility is necessary to ensure contraction, the rigid scaffold provided by the microtubules (as evidenced by high $G'$ values for $\Phi_A=0.25$ composites) is also necessary to sustain steady contraction and correlated restructuring of actin and microtubule networks over large spatiotemporal scales. At $c_M=0.24$ μM, when $\Phi_A$ is increased to 0.75, networks exhibit accelerating contractile dynamics (Figs. 5, 6) as well as substantial and decorrelated restructuring of actin and microtubule networks (Figs. 3, 4).

The significantly larger increase in $\lambda(t)$ for the microtubules compared to actin filaments in $\Phi_A=0.75$ composites (Fig. 3B), also seen in the simulated contractile contour maps (Fig. 4A), suggests that microtubules are aligning and bundling more so than actin filaments, likely driven by entropic depletion interactions from the surrounding higher concentration actin network and facilitated by the active contraction of the actin network. This bundling increases the mesh size of the microtubule network and decreases its connectivity with the actin network, thereby providing less of a reinforcing scaffold to rigidify the actin network[37,38]. The microtubules are thus no longer able to provide sufficient extensile support against accelerating contraction. Notably, the average velocity and acceleration we measure is comparable for both actin and microtubule dynamics despite the ~135% difference in correlation length changes during activity. This separability of time-evolving structure and dynamics, not previously reported in single-substrate active systems, highlights the novel properties that can emerge in composite systems, which may prove useful in filtration and drug delivery applications where controlled structural changes are necessary to dynamically release or sequester payloads or toxins.

At equimolar concentrations of actin and microtubules ($\Phi_A$=0.5), we find another emergent regime in which restructuring can be tuned independently of dynamics by varying myosin concentration. Specifically, at $\Phi_A$=0.5, the contraction rate of networks with $c_M$>0.12 µM stays constant with increasing myosin concentration, in line with previous studies of disordered actomyosin networks[17]. However, while the contraction rate stays constant, $\lambda(t)$ increased dramatically with increasing motor density. We can understand this interesting effect as due to increased actin bundling at higher $c_M$ as the motors pull and crosslink actin filaments. For a connected network with modest motor concentration, motors pull on individual filament pairs that collectively contract towards a central region. As $c_M$ increases, more motors bind to a single filament, pulling on filament pairs and causing more local bundling and aggregation. Rather than contributing to increasing the contraction rate, the added energy input from the increased motor density goes to restructuring the network from a homogeneous mesh of individual filaments to that of bundles and aggregates with larger mesh sizes. Specifically, at $c_M$=0.24 µM, the spacing between myosin minifilaments is ~6 µm, comparable to the ~7 µm length of actin filaments in the composite[27]. Thus, all filament pairs have ~1-2 bound motors. As $c_M$ is doubled to 0.48 µM, the number of bound motors also doubles to ~2-4 so there are more motors than can act on a filament pair or cluster to contract a local region. This interpretation aligns with previous results for crosslinked actin-microtubule composites that showed that increasing actin crosslinkers beyond a critical density resulted in restructuring of the actin network from a uniformly crosslinked filament mesh to a bundled network with larger voids[38]. These larger voids, in turn promoted microtubule mobility by providing less of a connected meshwork with which the microtubules could entangle to hinder their motion. This second order effect of actin bundling, namely increasing microtubule mobility, allows the microtubule network to restructure just as quickly as the actin network during contraction despite the more rigid nature of microtubules.

Our collective results, which comprehensively map the dynamics, structure and mechanics of myosin-driven actin-microtubule composites to the concentrations of the constituent proteins, demonstrate the promise that cytoskeleton composites hold for creating tunable and resilient active materials. While crosslinked actomyosin systems have been extensively studied as a model for adaptive biomaterials[14], here we show that coupling microtubules to actomyosin systems greatly expands the phase space of possible dynamics and structure that cytoskeleton-inspired materials can access and confers emergent and useful properties not attainable with single-substrate systems.

Notably, the active dynamics and restructuring we report are obtained in the absence of crosslinkers and at lower protein concentrations than required for robust actomyosin contraction. Our materials design goal is to identify the regions of phase space in which our materials exhibit sustained and steady ballistic contraction while maintaining network connectivity and resilience. However, the comprehensive phase map we present not only provides insight into the different routes the cytoskeleton may use to alter its dynamics and structure, but also serves as a blueprint for the design of active and adaptable flexible-stiff polymer materials that can be tuned for diverse applications including wound healing, micro-actuation, filtration, drug-delivery and soft robotics.

**Materials and Methods**

***Protein Preparation*:** The following cytoskeleton proteins are purchased: porcine brain tubulin (Cytoskeleton T240), HiLyte Fluor-488-labeled tubulin (Cytoskeleton TL488M-A), rabbit skeletal actin (Cytoskeleton AKL99), Alexa-568-labeled actin (ThermoFisher A12374), and rabbit skeletal myosin II (Cytoskeleton MY02). Both labeled and unlabeled tubulin are reconstituted to 5 mg/ml with 80 mM PIPES (pH 6.9), 2 mM $MgCl_2$, 0.5 mM EGTA, and 1 mM GTP. Unlabeled actin is reconstituted to 2 mg/ml in 5 mM Tris-HCl (pH 8.0), 0.2 mM $CaCl_2$, 0.2 mM ATP, 5% (w/v) sucrose, and 1% (w/v) dextran. Labeled actin is reconstituted to 1 mg/ml in 5 mM Tris (pH 8.1), 0.2 mM $CaCl_2$, 0.2 mM dithiothreitol (DTT), 0.2 mM ATP, and 10% (w/v) sucrose. Myosin II is reconstituted to 10 mg/ml in 25 mM PIPES (pH 7.0), 1.25 M KCl, 2.5% sucrose, 0.5% dextran, and 1 mM DTT. All cytoskeleton proteins are flash frozen in experimental-sized aliquots and stored at -80°C. Immediately prior to use, enzymatically dead myosin is removed from myosin II aliquots using a spin-down protocol previously described[29].

***Composite Network Assembly*:** To form actin-microtubule composite networks, actin monomers and tubulin dimers are polymerized for 30 min at 37°C in PEM-100 (100 mM PIPES, 2 mM $MgCl_2$, and 2 mM EGTA) supplemented with 0.1% Tween, 1 mM ATP, and 1 mM GTP. Actin and tubulin are added in concentrations varying from 1.4 – 4.4 μM such that the total protein concentration is maintained at 5.8 μM. Actin filaments are stabilized with the addition of an equimolar ratio of phalloidin to actin (1.4 – 4.4 μM), and microtubules are stabilized with the addition of a saturating concentration of Taxol (5 μM)[39,40]. To facilitate fluorescent imaging, 18% of actin monomers and 10% of tubulin dimers are labeled with Alexa-568 and HiLyte Fluor-488 conjugate dyes, respectively. Immediately prior to imaging, an oxygen scavenging system (45 µg/ml glucose, 0.005% β-mercaptoethanol, 43 µg/ml glucose oxidase, and 7 µg/ml catalase) is added to reduce photobleaching and 50 µM blebbistatin is added to inhibit myosin activity until blebbistatin is deactivated with 488 nm light. Spun-down myosin II is added at varying concentrations ($c_M = 0.12 – 0.48$ µM) to generate network activity.

***Sample Preparation***: Sample chambers holding ~10 µL of solution are formed by adhering a glass coverslip to a glass slide using two strips of heated parafilm (~70 µm thick). Samples are loaded into chambers immediately after polymerization and chambers are sealed with epoxy. To create hydrophobic surfaces that myosin will not adhere to, coverslips and glass slides are passivated with 2% dichlorodimethylsilane as previously described[29,41].

***Confocal Microscopy***: Networks are imaged using a Nikon A1R laser scanning confocal microscope with a 60× 1.4 NA objective (Nikon). Actin filaments are imaged using a 561 nm laser with 561 nm excitation and 595 nm emission filters. Microtubules are imaged using a 488 nm laser with 488 nm excitation and 525 nm emission filters. Myosin II motor activity is controlled by deactivating blebbistatin with 488 nm excitation. 256×256 pixel images (212 µm × 212 µm) are taken in the middle of the ~70 µm thick chamber for 45 min at 2.78 fps. Examples of time-series are shown in Movie S1, and representative images are shown in Fig. 1. Experiments are performed on 3-5 different replicates.

***Differential Dynamic Microscopy (DDM)***: We perform differential dynamic microscopy (DDM) on each of the 3-5 time-series for each condition using custom Python scripts[42]. DDM analysis determines how quickly density fluctuations decay between images separated by a given lag time $\Delta t$[35]. We compute variances by taking a fast Fourier transform of differences in image intensity at different lag times, and the resulting power spectrum is radially averaged to generate an image structure function $D(q, \Delta t)$. For systems that plateau at long lag times (see Fig. 2), we fit $D(q, \Delta t)$ to the following model: $D(q, \Delta t) = A(q) \times (1 - f(q, \Delta t)) + B(q)$ where $q$ is the wave vector, $A(q)$ is the amplitude, $B(q)$ is the background, and $f(q, \Delta t)$ is the intermediate scattering function, which contains the dynamics of the system. To determine the type of motion and the corresponding rate, we model the intermediate scattering function as an exponential: $f(q, \Delta t) =$

$e^{-(\frac{\Delta t}{\tau(q)})^{\gamma}}$ where $\tau(q)$ is the decay time and $\gamma$ is the scaling factor. If $\tau \sim \frac{1}{kq}$ then the system exhibits ballistic motion and $k$ represents the corresponding velocity.

We analyze time-series over the entire 45 min or until the network disappears from the field of view. To determine whether dynamics changed over the course of activity, we also perform DDM analysis over consecutive smaller time intervals using the same Python script used for the 45-minute time-series analysis. Specifically, we divide the 7500 frames of each 45-minute series into successive 6-minute segments containing 1000 frames each. By comparing image structure functions for different time intervals in the time-series, we determine how the active dynamics vary over the course of the activity. We identify network acceleration during contraction by an upward trend in computed velocities over the 45-min activation time (Fig. 3).

*Spatial Image Autocorrelation (SIA):* We perform spatial image autocorrelation (SIA) analysis on each of the 3-5 time-series for each condition using custom Python scripts[43]. SIA measures the correlation in intensity $g$ of two pixels in an image as a function of separation distance $r$[36]. We generate autocorrelation curves by taking the fast Fourier transform of an image at a given time, multiplying by its complex conjugate, and then applying an inverse Fourier transform and normalizing by intensity:

$$g(r) = \frac{F^{-1}(|F(I(r))|^2)}{[I(r)]^2}$$

The correlation length $\lambda$ is determined by fitting the decaying section of each autocorrelation curve to an exponential: $g(r) = Ae^{\frac{-r}{\lambda}}$. To quantify how the composites rearrange over time, correlation lengths $\lambda$ are computed at consecutive 6-minute intervals over the experimental time frame.

*Modeling and Simulations:* We combine rigidity percolation theory[25,44–46] with an active double-network model made of a stiff microtubule network interacting with an active semiflexible actomyosin network. The construction of this active rigidly percolating double-network (RPDN)[47] is described in the Supplementary Information (SI). The bonds in the two networks are uniformly and randomly removed according to two different probabilities, $1-p_1$ for the stiff (microtubule) network, and $1-p_2$ for the semiflexible (actin) network, where $0 < p_1, p_2 < 1$, and a continuous series of colinear bonds constitute a fiber. The values of $p_1$ and $p_2$ are obtained from experimental molar concentrations of tubulin and actin. The stretching moduli of the fibers in the microtubule and actin networks, $\alpha_1$ and $\alpha_2$, as well as their bending moduli, $\kappa_1$ and $\kappa_2$, are calculated from the known persistence lengths and cross-section diameters of microtubules and actin filaments (see SI). The contraction of bonds in the actomyosin network is incorporated via a parameter $\delta$ which leads to increasingly reduced rest lengths as the myosin concentration is increased and is obtained from the ratio of experimental myosin and actin concentrations (described in SI). The two networks interact via weak Hookean springs of spring constant $\alpha_3$. The energy cost of deforming the RPDN is given by $E = E_1(p_1, \kappa_1, \alpha_1) + E_2(p_2, \kappa_2, \alpha_2, \delta) + E(p_1, p_2, \alpha_3)$, where $E_1$ is the deformation energy of the microtubule network, $E_2$ is the deformation energy of the actomyosin network, and $E_3$ is the deformation energy of the bonds connecting the two networks (see SI for details). The energy $E$ is minimized using a multi-dimensional conjugate gradient (Polak-Ribiere) method as a function of the bond occupation probabilities, $p_1$ and $p_2$, and contraction parameter $\delta$ to obtain states of the active RPDN for different concentrations of actin, myosin, and microtubules. Simulations performed under 0.005% shear are used to obtain the mechanical modulus $G'$ (described in SI). Note that this model does not incorporate time evolution, filament bundling, breaking, or buckling.

**Figures and Captions**

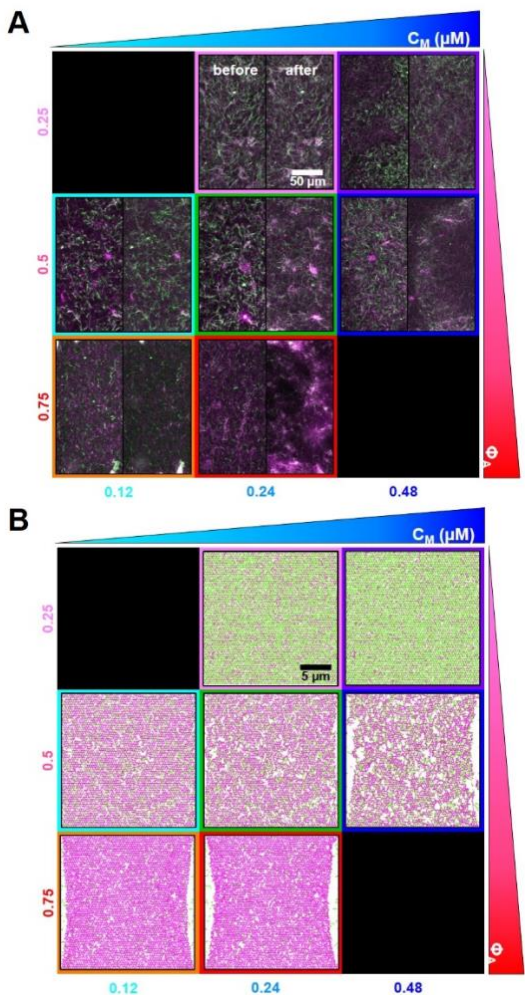

**Figure 1. Tuning the composition of active cytoskeleton composites.** **(A)** 256×128 pixel (212×106 µm) two-color confocal microscopy images show how composites of actin filaments (magenta) and microtubules (green) are rearranged via myosin II motor activity (Movie S1). In each panel, images taken at the beginning (left, before) and end (right, after) of the 45-min myosin activation are shown. Panels are ordered by increasing myosin concentration ($c_M$, blue) going from left to right and increasing molar fraction of actin ($\Phi_A$, red) going from top to bottom. The colors outlining each panel match the color coding used in subsequent figures. Scale bar pertains to all images. Panels in the top-left and bottom-right do not have data. **(B)** Deformation of a simulated active biopolymer double-network for parameters shown in (A). Simulations show how double-networks made of an actomyosin network (magenta) and a microtubule network (green) deform for varying concentrations of myosin and actin. As in (A), panels are ordered by increasing myosin concentration ($c_M$, blue) from left to right and increasing actin fraction ($\Phi_A$, red) going from top to bottom. Simulation box size is 24 ×20 µm for all images.

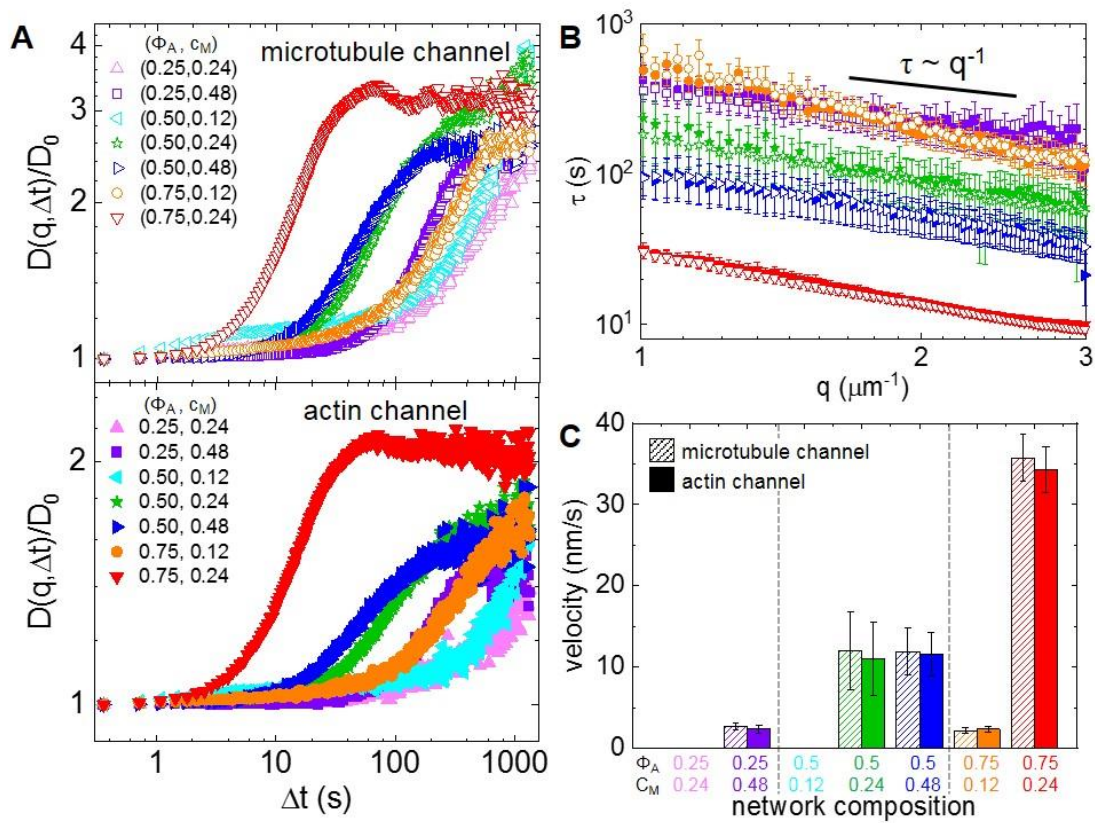

**Figure 2. Differential dynamic microscopy shows that composite contraction dynamics can be independently tuned by the actin fraction and myosin concentration.** (A) Representative image structure functions $D(q, \Delta t)$ of varying active composites, normalized by the corresponding initial value $D_0$, at wavenumber $q = 1.481$ µm$^{-1}$. DDM is performed separately on microtubule (top, open symbols) and actin (bottom, filled symbols) time-series channels. The lag-time dependence for all curves is largely indistinguishable when comparing actin and microtubule channels. For the actin channel, image structure functions of composites with lower myosin concentrations $c_M$ (cyan left pointing triangles) and lower actin fractions $\Phi_A$ (pink upwards pointing triangles) do not reach plateaus, while those of composites with higher myosin concentrations and actin fractions reach decorrelation plateaus at varying lag times (purple squares, green stars, blue right pointing triangles, orange circles, red downwards pointing triangles). (B) Average characteristic decay time $\tau$ vs wavenumber $q$ for both actin (filled symbols) and microtubule (open symbols) channels for each condition that reached a plateau in (A). All curves follow $\tau \sim q^{-1}$ scaling for $q = 1 - 3$ µm$^{-1}$, indicating ballistic motion. Symbol colors and shapes correspond to legends in (A). (C) Average contraction velocities $k$ extracted from fitting $\tau(q)$ curves in (B) to $\tau = (kq)^{-1}$. Error bars in (B) and (C) represent the standard error of values across all time-series.

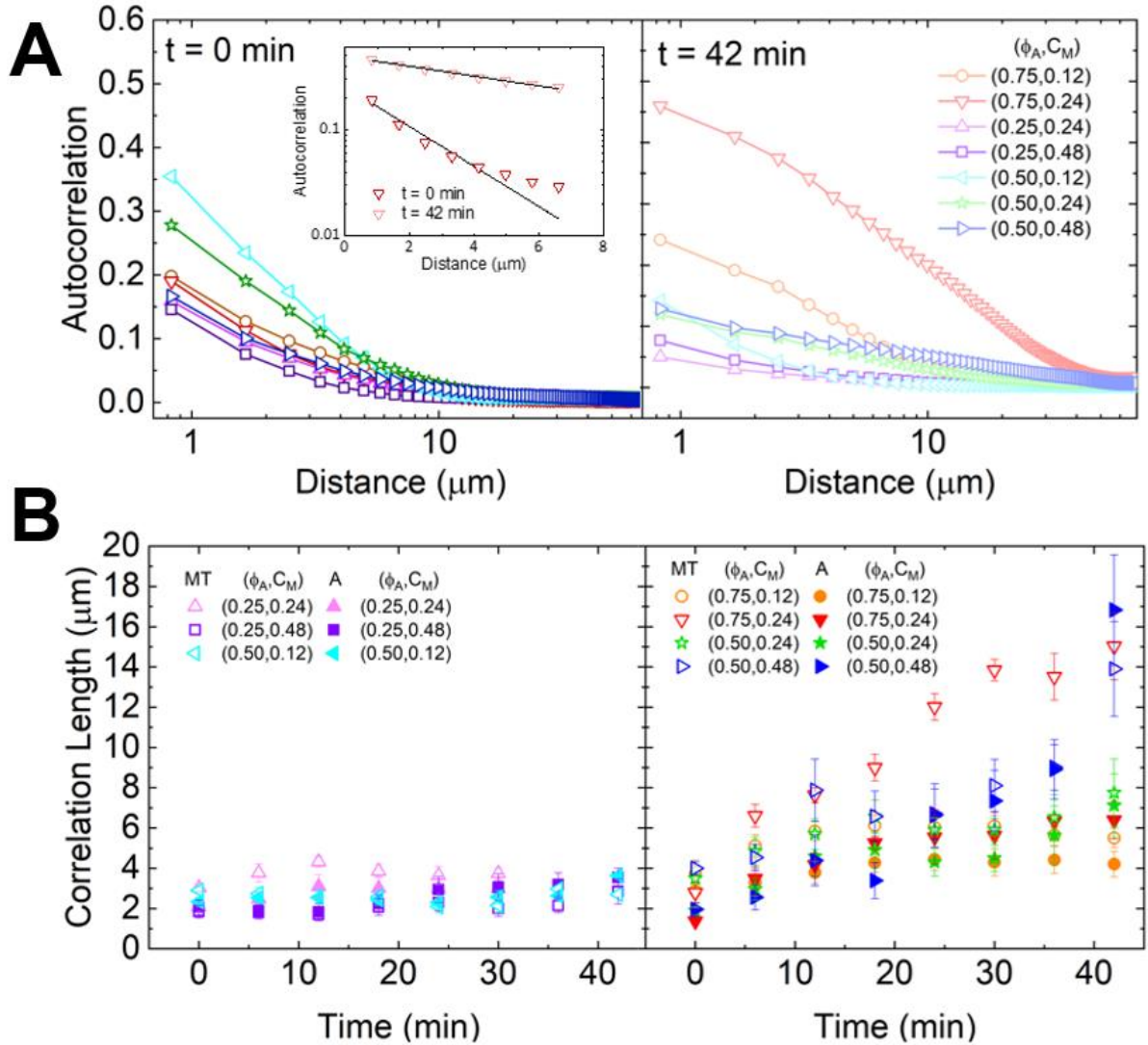

**Figure 3. Spatial image autocorrelation analysis reveals that myosin activity increases correlation lengths in a tunable manner independent of measured velocities.** (A) Average autocorrelation curves $g(r)$ for the microtubule channel for all composite formulations at the beginning (left, t = 0 min, dark shades) and end (right, t = 42 mins, light shades) of the experimental window. The inset shows example fits of $g(r)$ curves to $g(r) = Ae^{\frac{-r}{\lambda}}$ at the initial and final times for the $(\Phi_A, c_M) = (0.75, 0.12)$ composite. (B) Average correlation lengths for actin (closed symbols) and microtubule (open symbols) channels for each composite formulation determined from exponential fits (see inset in A) to the corresponding autocorrelation curves. Data is divided into composites that exhibit minimal restructuring (left) versus substantial restructuring (right). Error bars in (B) and (C) represent the standard error of values across all replicates.

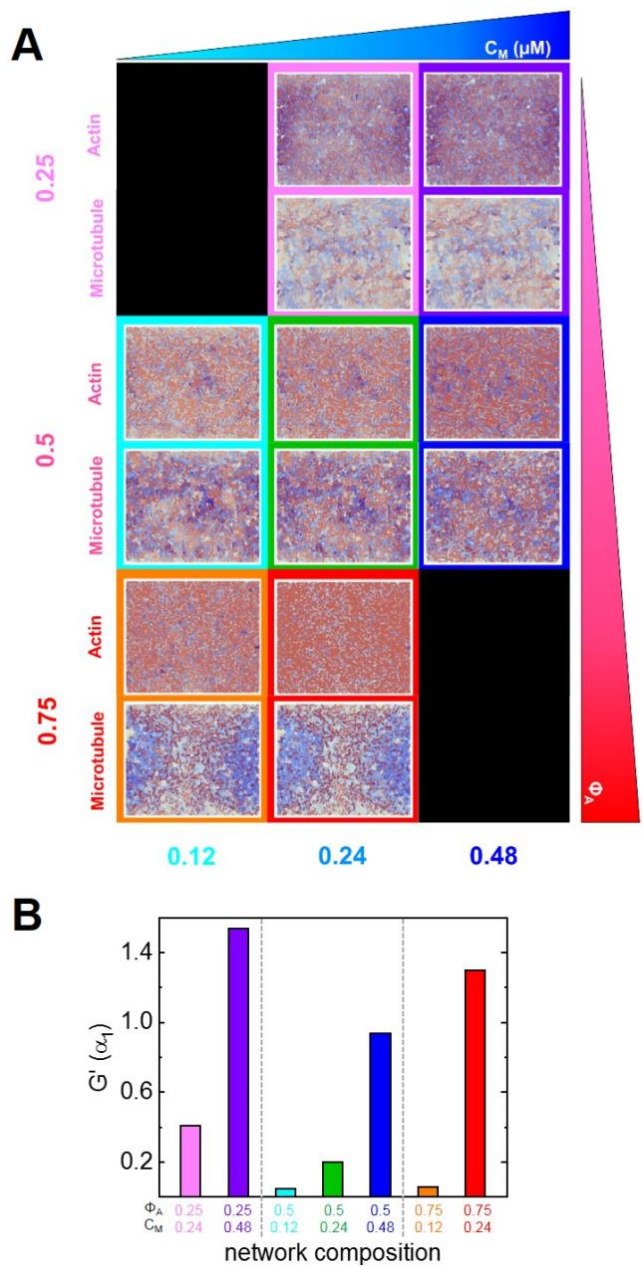

**Figure 4. Active double-network modeling shows tunable restructuring and rigidity of myosin-driven actin-microtubule composites.** (A) Heat maps for strains experienced by actin and microtubule networks in the active double-network model for the myosin concentrations and molar actin fractions shown in Fig. 1. Contractile strains are shown in shades of red and extensile strains are shown in shades of blue, with darker colors representing larger strains. White-colored regions represent negligibly small strains. The strain maps for the actin network and microtubule network for each condition are shown in the top and bottom of each panel, respectively. (B) Mechanical modulus $G'$ of a simulated active double-network for parameters shown in Fig. 1B. $G'$ is in simulation units corresponding to the microtubule stretching modulus ($\alpha_1$).

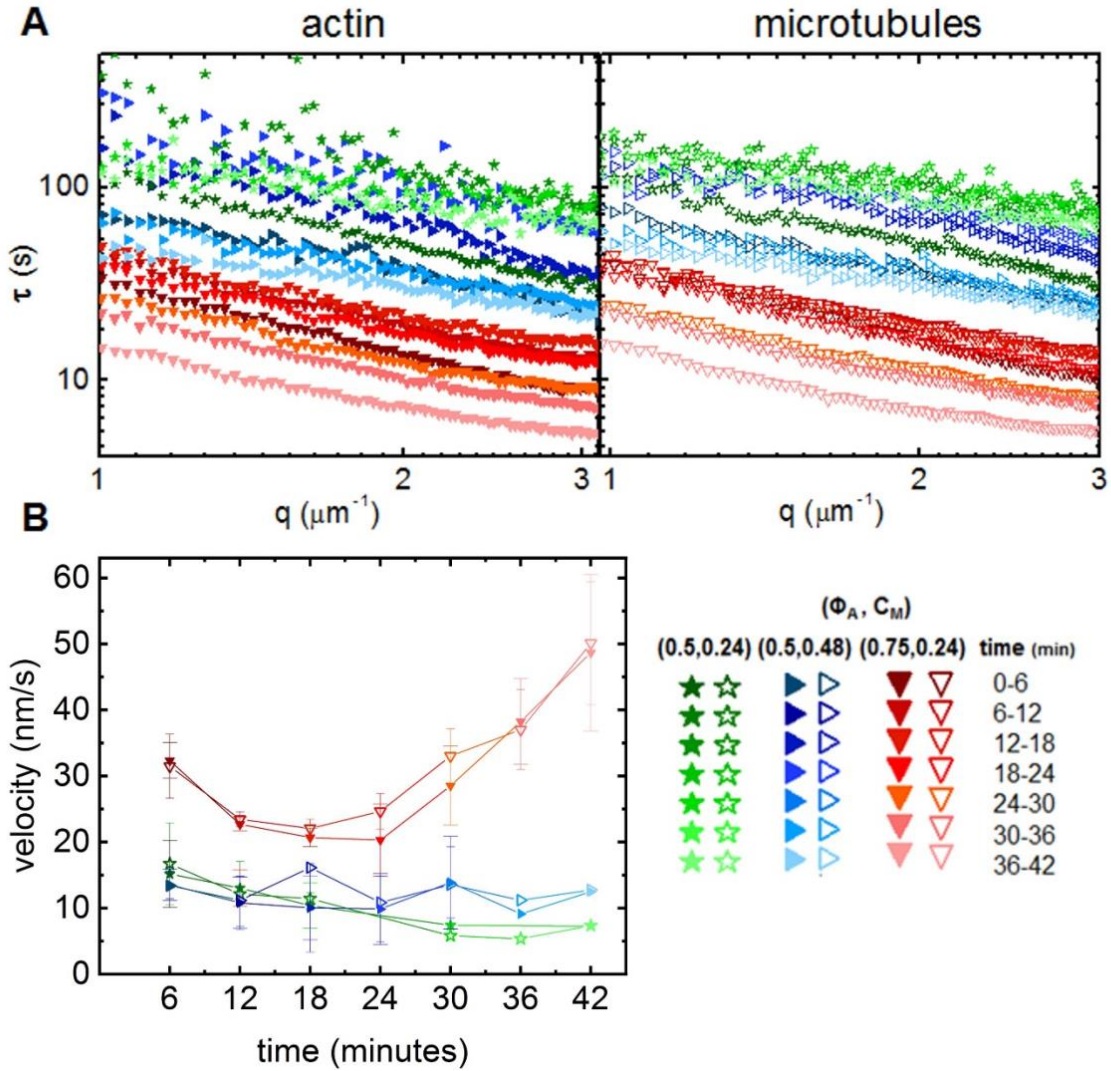

**Figure 5. Time-resolved differential dynamic microscopy shows higher actin fractions result in accelerating contraction dynamics.** (A) Characteristic decay time $\tau$ vs wavenumber $q$ for actin (filled symbols, left) and microtubule (open symbols, right) channels for consecutive 6-min intervals during the 45 min activation time of a representative time-series. Curves follow $\tau \sim q^{-1}$ scaling, indicating ballistic motion. Varying colors and symbols correspond to different composite formulations and time intervals as depicted in the legend (lower right). $\tau(q)$ curves for the $\Phi_A = 0.75$ composite (red, downward-pointing triangles) are lower in magnitude and show substantially more time-dependence compared to both $\Phi_A=0.5$ composites. Both $\Phi_A=0.5$ composites also have overlapping $\tau(q)$ curves, indicating negligible dependence of time-resolved dynamics on the myosin concentration $c_M$. (B) Contraction velocities for actin filaments (closed symbols) and microtubules (open symbols) for each 6-min interval of every analyzed time-series, extracted from fitting corresponding $\tau(q)$ curves. Error bars represent the standard error of values across all replicates. Symbol colors and shapes match Fig. 3 and correspond to ($\Phi_A$, $c_M$) combinations shown in the legend.

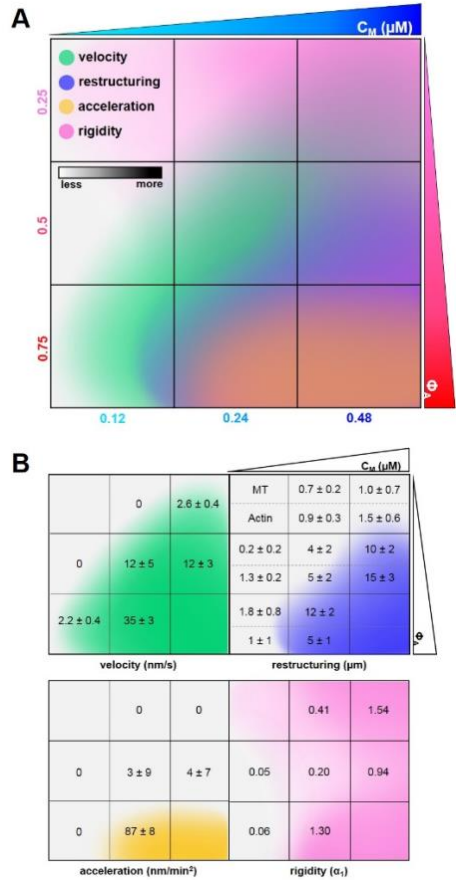

**Figure 6. Cartoon phase diagram of active cytoskeleton composite dynamics mapped to the corresponding composite formulation ($\Phi_A$, $c_M$) highlights tunability and desirable properties of active composites.** (A) Comprehensive phase diagram showing how myosin-driven composite velocities (green, determined from DDM), restructuring (blue, determined via SIA), acceleration (yellow, determined from time-resolved DDM), and rigidity (pink, determined via double-network simulations) vary in formulation space ($\Phi_A$, $c_M$). Darker shading of each color qualitatively indicates increasing metric values quantified within the separated figures in (B). Note that the panels in the bottom figures that lack numbers are those that we did not measure. The colors in those panels (top left, bottom right) are extrapolated based on the trends in our data. Velocities are averaged over both actin and microtubule channels as we observed similar dynamics for both filaments in all cases. Degree of restructuring is calculated by taking the difference in final and initial correlation length ($\Delta\lambda = \lambda_{final} - \lambda_{initial}$), and is reported separately for actin and microtubule channels because we observed instances of significant differences in $\Delta\lambda$ for the two filaments. Acceleration is calculated by computing the average change in velocity over time for the final 21 min of each time-series. Rigidity is calculated by performing simulations under 0.005% shear to obtain the mechanical modulus $G'$, expressed in simulation units corresponding to the microtubule stretching modulus $\alpha_1$. Composites with $\Phi_A=0.5$ and sufficiently high $c_M$ exhibit coordinated contractile dynamics and rearrangement without destructive network acceleration or decorrelation of actin and microtubule restructuring.

**Acknowledgements**

We thank S. Ricketts and B. Gurmessa for work in optimizing the polymerization and characterization protocols for actin-microtubule networks, L. Farhadi for sharing expertise on active actin-microtubule networks and spatial image analysis, S. Sahu for sharing expertise on coverslip passivation, V. Yadav and M. Murrell for sharing expertise on myosin II, and J. Garamella and K. Peddireddy for helpful discussion.

**Funding**

This research was funded by a William M. Keck Foundation Research Grant (awarded to R.M.R.-A., J.L.R., M.D., and M.J.R.), a National Institutes of Health R15 Award (National Institute of General Medical Sciences award no. R15GM123420, awarded to R.M.R.-A and R.J.M.), and a National Science Foundation Award (NSF Biomaterials award no. 1808026 to M.D.).

**Author Contributions**

R.M.R.-A. conceived the project, guided the experiments, interpreted the data, and wrote the manuscript. M.D. guided the simulations, interpreted the data, and helped write the manuscript. G. Lee performed the experiments, analyzed and interpreted the data, and wrote the manuscript. G. Leech performed the experiments, analyzed data, and helped write the manuscript. P.L. and J.M. performed the simulations, analyzed and interpreted the data, and helped write the manuscript. C. Currie analyzed data and helped write the manuscript. J.L.R., R.J.M., and M.J.R guided the experiments, interpreted the data, and provided useful feedback.

**Supplementary Information**

**Active Cytoskeletal Composites Display Emergent Tunable Contractility and Restructuring**

Gloria Lee[1], Gregor Leech[1], Pancy Lwin[2], Jonathan Michel[2], Christopher Currie[1], Michael J. Rust[3], Jennifer L. Ross[4], Ryan J. McGorty[1], Moumita Das[2], and Rae M. Robertson-Anderson[1]

[1]Department of Physics and Biophysics, University of San Diego
[2]School of Physics and Astronomy, Rochester Institute of Technology
[3]Department of Molecular Genetics and Cell Biology, University of Chicago
[4]Department of Physics, Syracuse University

**Contents:**

**Movie S1:** Sample time-series of active actin-microtubule composites corresponding to images shown in Figure 1 of manuscript.

**Figure S1:** Plateaus in DDM image structure functions are due to myosin-driven contractile dynamics.

**Figure S2:** Active entangled actin networks exhibit faster contraction dynamics than active actin-microtubule composites.

**Table S1:** Active composites exhibit ballistic contraction.

**Section S1:** Double-Network Model and calculations

**Figure S3:** Model of an active double network with molar actin fraction $\Phi_A = 0.5$. Actin filaments are shown in magenta and microtubules are shown in green.

**Figure S4:** A single primitive cell of a kagome fiber network.

**Section S2:** Bibliography

**Video S1. Time-series of active actin-microtubule networks show varying compositions tune dynamics and structure.** https://drive.google.com/file/d/1PlQFyRUjBFfkyA2vDHB_-k5fHO7othzW/view?usp=sharing

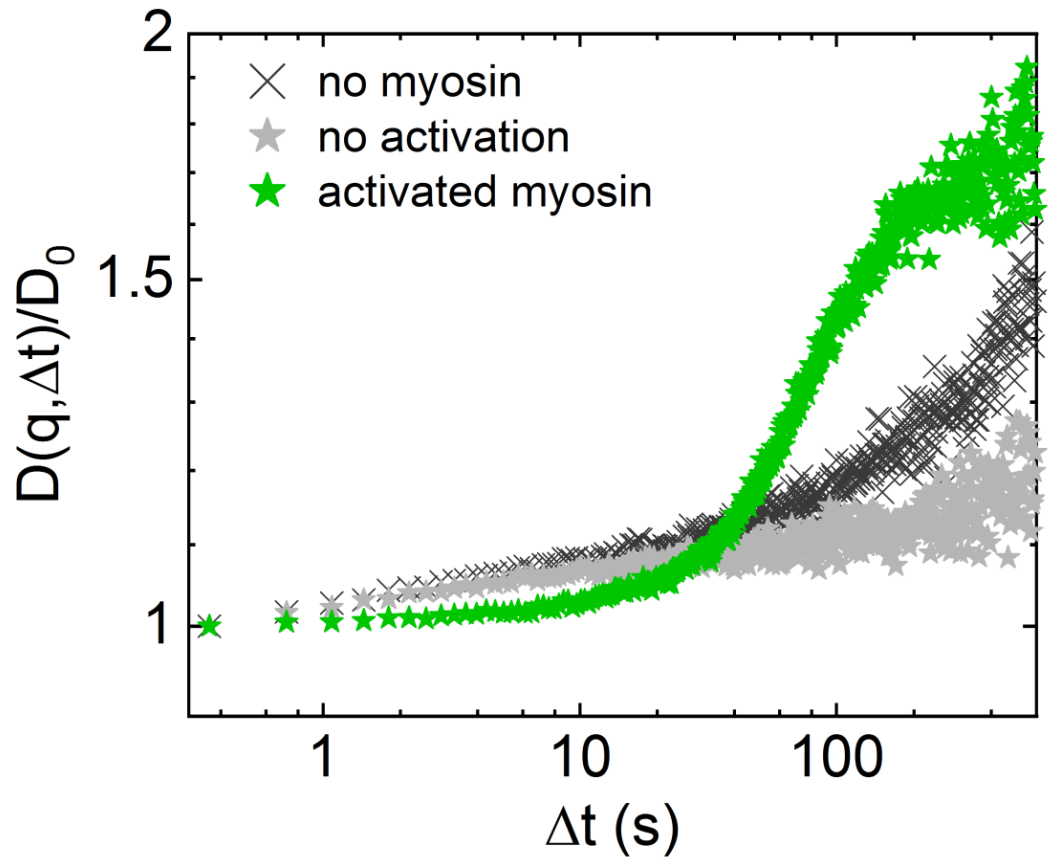

**Figure S1. Plateaus in DDM image structure functions are due to myosin-driven contractile dynamics.** Representative normalized image structure functions $\frac{D(q,\Delta t)}{D_0}$ at $q = 1.48$ μm$^{-1}$ in the actin channel for $(\Phi_A, c_M) = (0.50, 0.24)$ composites imaged under the following conditions: no myosin is included in the composite (dark grey crosses), myosin is included but the composite is not exposed to 488 nm light to deactivate blebbistatin (light grey stars), and myosin is included and the composite is exposed to 488nm light (green stars). Neither negative control case (light or dark grey) reaches a decorrelation plateau over the experimental time frame, demonstrating that image structure function plateaus are due to myosin-driven active dynamics.

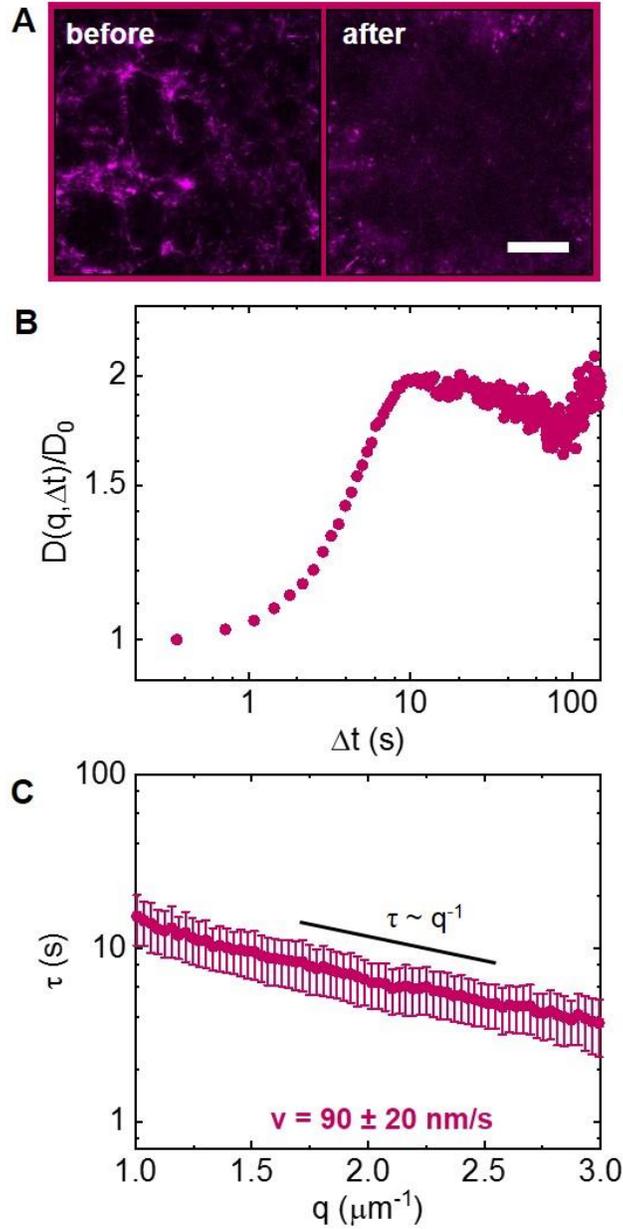

**Figure S2. Active entangled actin networks exhibit faster contraction dynamics than active actin-microtubule composites.** (A) Still images of a 5.8 μM actin network (($\Phi_A$, $c_M$)=(1, 0.24)) before and after 6 min of myosin II rearrangement. Timescale is shortened due to the network disappearing from the field of view due to faster contraction dynamics. Scale bar is 50 μm. (B) Representative normalized image structure function $\frac{D(q,\Delta t)}{D_0}$ at $q$ = 1.48 μm$^{-1}$. $\frac{D(q,\Delta t)}{D_0}$ reaches a plateau at a much earlier lag time compared to networks containing microtubules (see Fig. 2A). (C) Average $\tau(q)$ plot follows a power law relationship $\tau = \frac{1}{kq}$, from which we extract an average contraction velocity of 90 ± 20 nm/s, more than double the velocity of the ($\Phi_A$, $c_M$) = (0.75, 0.24) composite.

| ($\Phi_A$, $c_M$) | power (microtubule channel) | power (actin channel) |
|---|---|---|
| (0.25, 0.24) | -- | -- |
| (0.25, 0.48) | -0.97 ± 0.01 | -0.88 ± 0.08 |
| (0.50, 0.12) | -- | -- |
| (0.50, 0.24) | -1.00 ± 0.08 | -1.1 ± 0.1 |
| (0.50, 0.48) | -0.94 ± 0.05 | -0.87 ± 0.04 |
| (0.75, 0.12) | -1.5 ± 0.2 | -1.5 ± 0.1 |
| (0.75, 0.24) | -1.12 ± 0.07 | -1.10 ± 0.07 |

**Table S1. Active composites exhibit ballistic contraction.** For image structure functions that plateau, $\tau(q)$ plots are fit to the form $\tau \sim q^b$ over the range $q$=1-3 µm$^{-1}$, where $b$ is the power displayed in the table. Values are averaged over 3-5 time-series.

**Section S1: Double-Network Model and Calculations**

**S1a: Mathematical model for simulating myosin-driven active double-networks composed of actin filaments and microtubules**

Rigidity percolation theory has been immensely successful in predicting the mechanical properties and phase transitions in single-component cytoskeletal and extracellular matrix networks as a function of filament concentrations. This theory models biopolymer networks as two interconnected networks of disordered fibers and provides a framework for connecting network rigidity to structure and composition. Here we combine rigidity percolation theory with an active double network model made of a stiff microtubule network and an active semiflexible actomyosin network.

This active rigidly percolating double-network (RPDN) is constructed as follows. Starting with two networks, each based on a fully occupied kagome lattice such that at each crosslink there are no more than two crossing fibers, we dilute the networks by uniformly and randomly removing bonds from the networks according to two different probabilities (Fig. S3). We remove bonds from the stiff microtubule network with probability $1 - p_1$ and from the semiflexible actin network with probability $1 - p_2$, where $0 < p_1, p_2 < 1$, and a contiguous series of colinear bonds constitute a fiber. The stretching moduli of the fibers in the stiff and semiflexible networks are $\alpha_1$ and $\alpha_2$ respectively, and the bending moduli are $\kappa_1$ and $\kappa_2$ respectively. The two networks interact via weak Hookean springs with spring constant $\alpha_3$, which connect the midpoints of bonds $(x_1, x_2)$ and are only present when corresponding bonds are present in both networks. The energy cost of deforming this double network is given by:

$$E_1 = \frac{\alpha_1}{2} \sum_{<ij>} p_{1,ij} \left(r_{ij} - r_{ij0}\right)^2 + \frac{\kappa_1}{2} \sum_{<\widehat{ijk}=\pi>} p_{1,ij}\, p_{1,jk}\, \Delta\theta_{ijk}^2$$

$$E_2 = \frac{\alpha_2}{2} \sum_{<ij>} p_{2,ij} \left(s_{ij} - \rho s_{ij0}\right)^2 + \frac{\kappa_2}{2} \sum_{<\widehat{ijk}=\pi>} p_{2,ij}\, p_{2,jk}\, \Delta\beta_{ijk}^2$$

$$E_3 = \frac{\alpha_3}{2} \sum p_{1,ij}\, p_{2,ij} (x_1 - x_2)^2 \quad (1)$$

where $E_1$ is the deformation energy of the stiff network, $E_2$ is the deformation energy of the semiflexible network, and $E_3$ is the deformation energy of the bonds connecting the two networks. In $E_1$ and $E_2$, the first term corresponds to the energy cost of fiber stretching, and the second term to fiber bending[1].

In the above expression, the indices $i, j, k$ refer to sites (nodes) in each lattice-based network, such that $p_{ij}$ is 1 when a bond between those lattice sites is present and 0 if a bond is not present. The quantities $r_{ij}$ and $s_{ij}$ refer to the vector lengths between lattice sites $i$ and $j$ for the deformed stiff and flexible networks respectively, while $r_{ij0}$ and $s_{ij0}$ are the corresponding quantities for the initial undeformed networks. Active contractility is incorporated into the semiflexible network by setting the rest length of the bonds in this network to be $\rho s_{ij0}$, where $\rho$ is a function of myosin concentration as described later in this document and is 1 for a purely actin network and less than 1 for an actomyosin network[2]. The angles $\Delta\theta_{ijk}$ in the rigid network and $\Delta\beta_{ijk}$ in the semiflexible network correspond to the change in angles between initially collinear bond pairs $ij$ and $jk$ for the deformed and undeformed network, respectively.

Simulations of the above active RPDN model determined the linear response under 0.005% shear. We adopt a shear protocol where external deformations are applied along the top and bottom boundaries and periodic boundary conditions are used for the left and right sides of the network. For each set of parameters, active double networks containing $\sim 2 \times 10^5$ nodes were randomly generated with given fractions of bonds $1 - p_1$ and $1 - p_2$ missing. The total deformation energy was minimized for the applied macroscopic shear and the shear modulus was calculated[1] as a function of the bond occupation probabilities $p_1$ and $p_2$. The values of $p_1$ and $p_2$ were obtained from experimental molar concentrations of tubulin and actin respectively, and the contraction parameter $\rho$ was obtained using the ratio of experimental myosin and actin concentrations as described in the next section.

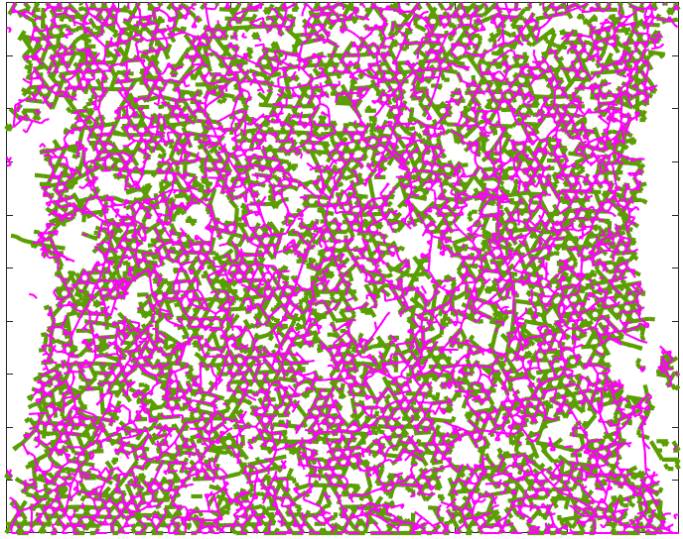

**Figure S3:** Model of an active double-network with molar actin fraction $\Phi_A$=0.5. Actin filaments are shown in magenta and microtubules are shown in green.

## S1b: Parameter estimation informed from experiments and literature

Mesh sizes are calculated for varying composites using the equations $\eta_A = \frac{0.3}{\sqrt{c_A}}$ and $\eta_{MT} = \frac{0.89}{\sqrt{c_T}}$, which relate mesh size to polymer concentrations for actin and tubulin respectively[3]. To calculate stretching moduli and bending moduli, we use persistence lengths of 20 μm and 1 mm, and diameters of 10 nm and 50 nm, for the actin filaments and microtubules respectively[3].

By defining persistence length $l_p$ as $l_p = \frac{\kappa}{k_B T}$, where $\kappa$ is the bending stiffness of the bond, $k_B$ is the Boltzmann constant and $T$ is the temperature, the bending modulus is given by the relation $\kappa \propto l_p$. For slender rods, the stretching modulus is given by $\alpha \propto \frac{l_p}{R^2}$, where $R$ is the cross-sectional radius of the rod [4]. From these two relationships, we can calculate the stretching moduli $\alpha_1$ and $\alpha_2$, and the bending moduli $\kappa_1$ and $\kappa_2$, of the fibers in the stiff and semiflexible networks respectively. In our simulations, all moduli are scaled by, and expressed in terms of, $\alpha_1$. To introduce contraction to the actomyosin network, we assign different values of $\rho$, or the amount the rest length of each bond is reduced due of myosin-induced contractility[2], to networks with different myosin concentrations. We use the Fermi estimate $\rho = 1 - \frac{[myosin]}{[actin]}$, such that in the absence of myosin, the network does not undergo any contraction.

To map the concentrations of actin filaments and microtubules to the bond occupation probabilities in the semiflexible and rigid networks, we use the following procedure. Let each bond in the network have length $l_0$, and let a primitive cell of the network have a length $a = l_0$, as shown in Fig. S4.

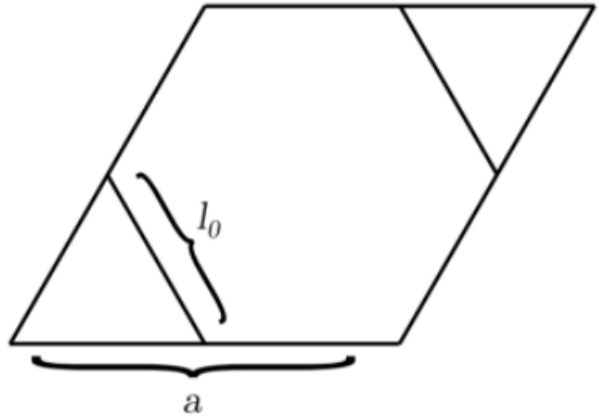

**Figure S4:** A single primitive cell of a kagome fiber network

There are eight perimeter bonds, each contributing half a bond to the primitive cell, and two interior bonds, each contributing one bond. The area of a primitive cell is $A = \frac{\sqrt{3}a^2}{2} = 2\sqrt{3}l_0^2$. For a bond occupation probability $p$, the length of fiber per unit area is then $\frac{length}{area} = \frac{6pl_0}{2\sqrt{3}l_0^2} = \frac{p\sqrt{3}}{l_0}$. Assuming the thickness of each slice of the three-dimensional sample to be $l_0$, the length per unit volume is then $\frac{length}{volume} = \frac{p\sqrt{3}}{l_0^2}$. For a given fiber type $f$, let $\lambda_f$ be the number of monomers per unit length. If the molar concentration is $[f]$, then the total fiber length per unit volume should be $\frac{length}{volume} = \frac{[f]N_A}{\lambda_f}$ where $N_A$ is Avagadro's constant $\approx 6.022 \times 10^{23}$. By setting the two length per volume equations equal to each other, $\frac{[f]N_A}{\lambda_f} = \frac{p\sqrt{3}}{l_0^2}$, we find $p = \frac{l_0^2[f]N_A}{\lambda_f\sqrt{3}}$.

To determine $\lambda_f(actin)$ and $\lambda_f(microtubule)$, which represent the number of monomers per unit length for actin filaments and microtubules, respectively, we use 2.7 nm per monomer for actin filaments[5] and 12 nm per 13 tubulin dimers for microtubules[6]. In using the above-mentioned calculation to map the microtubule and actin fractions to the bond occupation probabilities $p_1$ and $p_2$ respectively, we have set each bond occupation probability to 1 when the corresponding concentration is 5.8 μM.

**S1c: Creating strain maps of actin and microtubule networks**

To produce strain maps, we begin by producing a Delaunay triangulation[7] of the vertices of the network in its undeformed state and compute the area of each triangular facet. After the bonds of the actin network are subjected to contractile forces, and the network is relaxed to its energetic ground state, we use the new positions of the vertices in the microtubule and actin networks to produce deformed triangular meshes. We compute the new area of each deformed triangular mesh cell, subtract from this area the area of the corresponding, undeformed mesh cell, and divide by the area of the undeformed cell to find relative change in area. We next identify all groups of contiguous cells that undergo contraction, and all contiguous groups that undergo extension. To color-code contiguous regions that undergo contractile or extensile deformation, we take the fourth root of each relative extension and contraction to account for the large dynamic range in contraction and extension. We map these adjusted values to a color gradient in which deep red regions are highly contractile, and deep blue values are highly extensile.

**Section S2: Bibliography**